\documentclass[letterpaper,10pt]{article} 

\usepackage{osameet3} 

\usepackage{amsmath,amssymb}
\usepackage[colorlinks=true,bookmarks=false,citecolor=blue,urlcolor=blue]{hyperref} 
\usepackage{cite,graphicx,amsmath,psfrag,bm}
\usepackage{subfigure}

\begin{document}

\title{Compact grating coupler using asymmetric waveguide scatterers}

\author{Ashutosh Patri\textsuperscript{1,*}, Xiao Jia\textsuperscript{1}, Muhammad Mohsin\textsuperscript{2,3},\\ St\'ephane K\'ena-Cohen\textsuperscript{2}, and Christophe Caloz\textsuperscript{1}}
\address{\textsuperscript{1}Department of Electrical Engineering, Polytechnique Montr\'eal, QC H3T 1J4, Canada \\
\textsuperscript{2}Department of Engineering Physics, Polytechnique Montr\'eal, QC H3T 1J4, Canada\\
\textsuperscript{3}Currently with the National Research Council Canada, Ottawa, ON K1A 0R6, Canada}
\email{\textsuperscript{*}ashutosh.patri@polymtl.ca}
\copyrightyear{2019}

\begin{abstract}
We demonstrate a novel grating coupler design based on double asymmetric and vertically oriented waveguide scatterers to efficiently couple normally incident light to a fundamental mode silicon waveguide laying on a buried oxide layer. 
\end{abstract}


\section{Introduction}
Grating couplers are ubiquitous in integrated photonics for the conversion of free-space propagating light to guided light~\cite{taillaert2006grating}. To achieve unidirectional propagation, grating couplers are typically designed for obliquely incident light. This strategy, however, can lead to alignment difficulty and poor stability. For this reason, there have been significant efforts to design efficient grating couplers that work at normal incidence\cite{sarathy1994normal}. Typical coupling efficiencies for such couplers are much lower than their oblique angle counterparts, and limited \emph{a priori} by their lack of directionality.

There are different strategies to design normal-incidence grating couplers. The simplest one is to use a momentum-matched binary grating. However, such a structure wastes power both in the undesired direction of the waveguide and in the zeroth-order beams. To achieve unidrectionality, one can use a blazed grating with asymmetric saw-tooth profile. This suppresses the undesired coupled mode, but involves complex fabrication. A closely related strategy involves modulating the width of the binary grating elements to achieve unidrectionality. Each individual element within a diffraction-period of a binary-blazed grating works like a waveguide that provides phase matching by proapagtional phase delay~\cite{lalanne1999waveguiding}. However, due to small diffraction-period size, the coupling between these waveguiding elements does not allow for an appropriate phase-match. Moreover, such designs still suffers from loss due to the presence of zeroth diffraction order~\cite{yang2011high}. To reduce such loss, one may use Bragg reflectors or metallic mirrors below the waveguide structure, but this requires additional layers and may be incompatible with the overall fabrication process ~\cite{taillaert2002out}.

Here, we propose an alternative approach where the multiple waveguiding elements of a binary-blazed grating is replaced by a single asymmetric waveguiding structure to provide appropriate phase-match. Such a dielectric grating coupler could solve the aforementioned issues: 1)~it uses double -- horizontal \emph{and} vertical -- symmetry breaking, and hence suppresses all of the undesired diffraction orders; 2)~the diffraction period consists of a single waveguiding element that avoids the inter-waveguide coupling issue of binary-blazed gratings; 3)~it is composed of purely dielectric material, and hence has negligible absorption loss. As the overall grating coupler structure utilizes two different types of waveguides; vertically oriented waveguides as its diffractive elements and a horizontal waveguide to which the free-space power will be coupled into, we call them \emph{vertical} waveguide and \emph{slab} waveguide, respectively to avoid any confusion.

\section{Design Rationale}
We designed a single-mode silicon slab waveguide on a buried oxide layer (BOX) based for operation at 1550~nm. The slab waveguide operates in the fundamental transverse electric mode. Applying the grating equation, one finds a period required for coupling to normally-incident light, $\Lambda=\lambda/n_\text{eff}$, to be $\sim$600~nm, where $n_\text{eff}$ is the effective refractive index obtained from the slab waveguide dispersion relation.

By applying the reciprocity principle, the grating coupler problem can be transformed into the equivalent, but simpler analysis of a grating decoupler.  First, the slab waveguide coupled to the grating should be impedance-matched only at one of its ends, and fully reflective at its other end, which requires symmetry breaking in the horizontal plane. Second, the grating should decouple light only toward the top, which requires symmetry breaking in the vertical direction.

To break the horizontal-plane symmetry, we designed a $\pi$-shaped silicon vertical waveguide scatterer. Due to the presence of two different dielectric media at both sides of the  vertical waveguide, air at the top and silicon-BOX at the bottom, the overall scattering of the waveguiding element is also vertically asymmetric, and hence radiates with different phases toward the top and the bottom. However, this does not resolve the issue of the zeroth-order beam, since all the scatterer radiate with same phase toward the bottom. In addition, these scatterers, when coupled to the slab waveguide, radiate more effectively toward the bottom than toward the top due to a higher field concentration in the substrate as compared to air. To resolve these two issues, we inserted a cylindrical hole in every two $\pi$-scatterer in such a fashion that the phases radiated toward the bottom by the holey and hole-less scatterers are out-of-phase; as a result the radiation toward the bottom is significantly suppressed. The combined structure of both kind of $\pi$-scatterers is shown in Fig. 1a.

\begin{figure}[htbp]
  \centering
  \includegraphics[width=15.5cm]{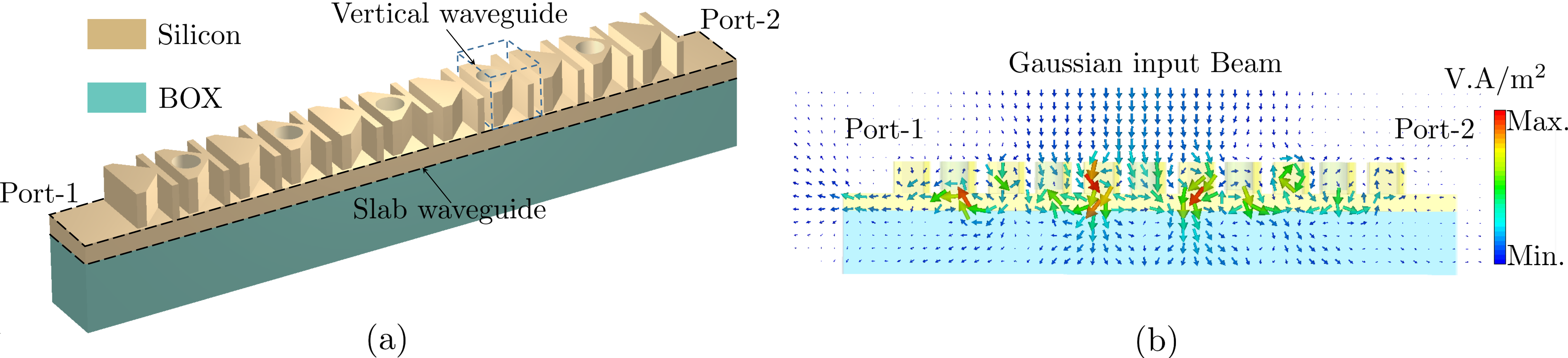}
\caption{Proposed asymmetric dielectric-scatterer grating coupler (a) Structure, (b) Poynting vector plot for Gaussian beam input.}
\end{figure}

\vspace{-0.5cm}

\section{Results}
We then performed a full-wave simulation of the grating coupler using periodic boundary conditions in CST Studio. We chose a 6.5 $\mu$m long grating coupler with 11 dielectric vertical waveguide scatterers to demonstrate a compact coupler design. In our reciprocal analysis, the slab waveguide reflects 7\% power from the desired end (Port-1), whereas it reflects 75\% power when fed from the undesired end (Port-2). In addition, when fed from the desired end, 10\% of the power is transmitted to the undesired end of the slab waveguide. This can be minimized to an almost negligible amount by choosing a longer coupler. The remaining 93\% power coming in from the desired end of slab waveguide splits into 2.5:1 between the radiation toward the top and the bottom. It should be noted that before introducing cylindrical holes in every two $\pi$-scatterers, the splitting ratio was 1:1.8 suggesting higher field concentration in the substrate.

Finally, to calculate the coupling efficiency, we irradiated the grating coupler with a Gaussian input beam, as shown in figure 1b, with a 2.5 $\mu$m radius spot size centered on grating coupler. Our simulation results show 40\% in-coupling efficiency to the desired waveguide direction. We should highlight that the radiation field profile during the inverse design is not matched to a that of a Gaussian input beam. Consequently, the overall efficiency can be further increased by laterally adjusting the grating elements to match the desired free-space profile. 



\begin{thebibliography}{99} 


\bibitem{taillaert2006grating} D. Taillaert, F. V. Laere, M. Ayre, W. Bogaerts, D. V. Thourhout, P. Bienstman, and R. Baets,``Grating couplers for coupling between optical fibers and nanophotonic waveguides,'' Jpn. J. Appl. Phys. \textbf{45}(8A), 6071--6077 (2006).

\bibitem{sarathy1994normal} J. Sarathy, R. A. Mayer, K. Jung, S. Unnikrishnan,D. L. Kwong, and J. C. Campbell, ``Normal-incidence grating couplers in Ge-Si,'' 	Opt. Lett. \textbf{19}(11), 798--800 (1994).

\bibitem{lalanne1999waveguiding} P. Lalanne, ``Waveguiding in blazed-binary diffractive elements," J. Opt. Soc. Am. A \textbf{16}(10), 2517--2520 (1999).

\bibitem{yang2011high} J. Yang, Z. Zhou, H. Jia, X. Zhang, and S. Qin, ``A High-performance and compact binary blazed grating coupler based on an asymmetric subgrating structure and vertical coupling,'' Opt. Lett. \textbf{36}(14), 2614--2617 (2011).

\bibitem{taillaert2002out} D. Taillaert, W. Bogaerts, P. Bienstman, T. F. Krauss, P. Van Daele, I. Moerman, S. Verstuyft, K. De Mesel, and R. Baets, ``An out-of-plane grating coupler for efficient butt-coupling between compact planar waveguides and single-mode fibers,'' IEEE J. Quantum Electron. \textbf{38}(7), 949--955 (2002).





\end{thebibliography}
\end{document}